\documentstyle[11pt,paspconf]{article}
\input psfig
\input epsf

\def\edcomment#1{\iffalse\marginpar{\raggedright\sl#1\/}\else\relax\fi}
\reversemarginpar

\def\lsim{\mathrel{\lower2.5pt\vbox{\lineskip=0pt\baselineskip=0pt
           \hbox{$<$}\hbox{$\sim$}}}}
\def\gsim{\mathrel{\lower2.5pt\vbox{\lineskip=0pt\baselineskip=0pt
           \hbox{$>$}\hbox{$\sim$}}}}

\newcommand{\cidfig}[6]{
    \protect\centerline{
    \epsfxsize=#1\epsffile[#2 #3 #4 #5]{#6}
}}

\def\q0{\hbox{q$_{\mbox{\scriptsize 0}}$}}
\def\H0{\hbox{H$_{\mbox{\scriptsize 0}}$}}
\def\Ho50{\hbox{H$_{\mbox{\scriptsize 0}} = 50$~\uniHo \ }}
\def\qo0p5{\hbox{q$_{\mbox{\scriptsize 0}} = 0.5$}}
\def\L0{\hbox{$\Lambda = 0$}}
\def\uniHo{\hbox{Km s$^{-1}$ Mpc$^{-1}$}}
\def\Msun{\hbox{M$_{\odot}$}}
\newcommand{\ldo}[1]{\hbox{$\lambda$#1 \AA}}
\begin{document}
\title{The colours of $z\approx2$ QSO host galaxies}
\author{Itziar Aretxaga$^{1}$, Roberto
J. Terlevich$^{2}$, B.J. Boyle$^{3}$}
\affil{$^{1}$ Max-Planck Institut f\"ur Astrophysik, Garching, Germany\\
$^{2}$ Royal Greenwich Observatory, Cambridge, UK\\
$^{3}$ Anglo-Australian Observatory, Epping, Australia}

\begin{abstract}
Three  high-redshift ($z \approx 2$) and
high luminosity ($M_B \lsim -28$~mag for \H0 = 50\uniHo, \qo0p5)
QSOs, two radio-quiet one radio-loud, 
were imaged in $R$, $I$ and $K$ bands. The luminosities,
colours and sizes of the hosts overlap with those of actively
star-forming galaxies in the local Universe.  
These properties give support to the young host galaxy interpretation
of the extended light around QSOs at
high-redshift.  The rest-frame UV and UV-optical colours are
inconsistent with the hypothesis of a scattered halo of light from the
active nucleus by a simple optically-thin scattering process produced
by dust or hot electrons.  If the UV light is indeed stellar, star
formation rates of hundreds of solar masses per year are implied, an
order of magnitude larger than in field galaxies at similar redshifts and
above.  This might indicate that the QSO phenomenon 
is preferentially accompanied by enhanced galactic activity at high-redshift.
\end{abstract}

\section{Introduction}
Several attempts to find the early
counterparts of big spheroids in large areas of the sky at high
redshifts have been
conducted but to date just a handful
of candidates  have been found [10].
% (Pritchet 1994, PASP, 106, 1052). 
Where are these bright objects hiding? 
In galaxy formation models based on the hierarchical merging of dark
matter halos the majority of large spheroids are formed at lower
redshifts,
%(Kauffmann \& Charlot 1997, MNRAS, in press)
 so that the
present-day large systems were previously broken-up into small fainter
pieces of a yet not completely assembled whole [8].
There is also a good possibility that some large spheroids
may have gone undetected as the host galaxies of bright QSOs.  Two
pieces of observational evidence support this hypothesis.  First, a
significant number of nearby bright QSOs are associated with luminous
ellipticals [3].
%(e.g. Bahcall et al. 1997, ApJ, 479, 642 ). 
Secondly, many of the properties of 
high-$z$ QSOs are
consistent with their association with the cores of early luminous
spheroids. Among them,  the high metal content of the Broad Line
Region of high-$z$ QSOs [6];
%(Hamman \& Ferland 1993, ApJ, 418, 11); 
 the detection of
large masses of dust in $z\approx 4$ QSOs [7];
%(Isaak et al. 1994, MNRAS,269, 28); 
and 
the correspondence of the observed luminosity function of QSOs and
ellipticals [11],
%(Terlevich \& Boyle 1993, MNRAS, 262, 491) 
do
suggest that QSO activity might be occurring in the central regions of
young massive ellipticals.

All these points are reinforced by the recent discovery that about 5
to 20\% of the luminosity of both radio-loud and radio-quiet QSOs at
$z \approx 2$ arises from extended structures of FWHM$\,\approx
1-2$~arcsec [1,2,9].
Unraveling the nature of the extensions around QSOs
at these redshifts is of major importance since they could be
revealing the hidden signature of the elusive early large spheroid
population.

In this contribution we focus our attention on the multicolour properties of
the hosts of three $z \approx 2$ QSOs, in an attempt to characterize
the emission mechanism.

\section{Data analysis}
The three QSOs, two radio-quiet 
1630.5$+$3749 and Q~2244$-$0105, one radio-loud PKS~2134$+$008,
were observed 
in the Harris $R$ and $I$ passbands 
with the auxiliary port of the 4.2m William Herschel Telescope (WHT) 
at the Observatorio de Roque de los Muchachos in La Palma, and in $K$
band with the Cassegrain focus of the 3.5m at the Observatorio de Calar 
Alto in Almer\'{\i}a (Spain). 
The QSOs were
selected so as to have bright stars in the field
($20 \lsim \theta \lsim 50$~arcsec), which enabled us to 
define the point spread function (PSF)
of each observation accurately (see [1,2] for 
a detailed description of data and analysis). 

For each QSO field, we 
defined a PSF using the brightest of the closest stellar 
companion to the QSO.
We then subtracted a scaled version of the
PSF in order either 
a) to produce zero counts in the center of the residuals
or b) to achieve a flat-top 
profile with no depression in the center.
We regard 
these quantities as lower limits (3--7\% of the QSO luminosity)
and best estimates (6--18\% of the QSO luminosity), respectively, of 
the total luminosity
of these extended components. 
We have detected $R$ and $I$-band residuals revealing
hosts in the three QSOs.
In all cases, the FWHM of the flat-top residual profiles ($0''.8$ to
$1''.1$) are significantly larger than the FWHM of the stars in each field
(typically 0.7'') -- see in Figure 1 the residuals left in our best
detection case, 1630.5$+$3749.

\vspace*{0.2cm}
%\begin{figure*}
    \cidfig{5.5in}{103}{595}{544}{747}{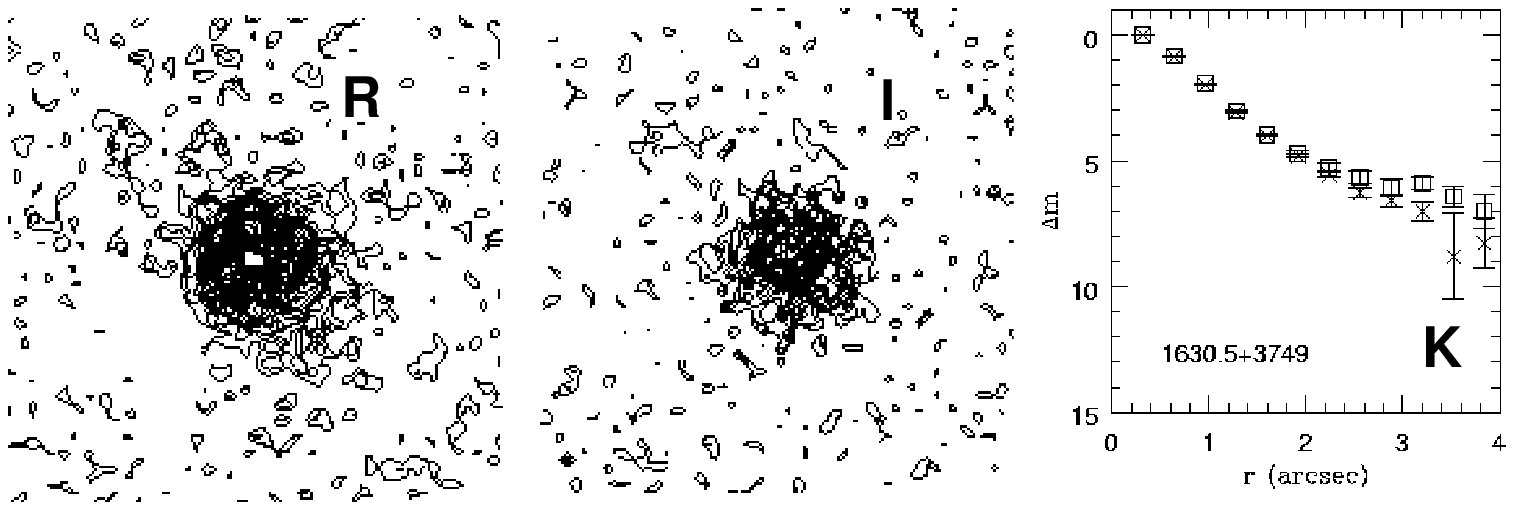}
%    \caption

%\vspace*{-0.3cm}
\noindent
{Figure 1: Countour plots of $R$ and $I$-band residuals after 
performing a PSF subtraction that leaves zero
counts in the center; and $K$-band radial profiles of QSO (squares)  
and star (crosses), normalized to their central flux.}
%\end{figure*}

\section{The nature of the hosts}

The hosts we have detected have luminosities that contribute between 5
and 12\% of the luminosity of the QSOs (nucleus$+$extension) both in
$R$ and $I$-bands (\ldo{2300} and \ldo{2800} rest-frame).  If the
excess over the stellar $K$-profile of 1630.5$+$3749 at $r\approx3''$ 
is real, the
$K$-band (\ldo{7600} rest-frame) luminosity of the host as derived
from the measured colours would also contribute to the total $K$-band
QSO luminosity about 5\%.  Our measurements indicate large and
luminous extended systems ($D_{eff}\approx 4$~Kpc, $R\approx21$ to 22~mag).  
We will consider two alternative
explanations for their nature: scattered light from the active
nucleus and stellar light.

\subsection{Scattering}
Large structures detected in high-redshift radio-galaxies
have been attributed to light scattered 
from the active nucleus by a powerful transverse radio-jet.  
Indeed, recent spectro-polarimetry of two
$z\approx1$ radio-galaxies that exhibit 'alignment effects' (optical
structures oriented in the direction of the radio-jet) indicates that
around 80\%\ of the total UV continuum emission at rest-frame
\ldo{2800} is non-stellar scattered light [5].
%(Cimatti et al. 1997,ApJ, 476, 677).
Since radio-galaxies and radio-loud QSOs could be identical objects
viewed from different angles, an important amount of
scattered light may be present around radio-loud nuclei.

The host of the core-dominated radio-loud QSO studied in this paper,
PKS 2134$+$008, has an $R-I$ colour that is 0.52~mag redder than the
nucleus. This argues against the simple optically-thin scattering
case, which would yield colours as blue as or bluer than the nucleus
itself. 
 The $R-K$ limits for this host ($\lsim 3.3$~mag) are not deep
enough to probe the optically-thin scattering case value.  Note that
four lobe-dominated radio-loud QSOs at $z\approx 2-2.5$
[9] have hosts with $B-K$ colours redder
than the nuclei.  However the $B-K$ (or $R-K$)
analysis discriminates poorly the
origin of the light below the \ldo{4000} break, since scattered 
light would contribute predominantly in $B$-band (rest frame
\ldo{1400}) 
and stellar light  would contribute predominantly in  $K$-band
(rest-frame \ldo{7300}).

The host of our radio-quiet QSO 
1630.5$+$3749, exhibits the same general properties, with colours 0.25
and 0.8~mag redder than the nucleus in $R-I$ and $R-K$, respectively.
These colours
are inconsistent with the optically-thin scattering case as an
explanation for the UV hosts of our QSOs. An optically-thick medium
should be invoked in order to produce colours redder than those of the
scattered source, but then the geometry of the scatterers should not
be symmetrical since we still see the blue colours of the nucleus
itself. This could be an alternative mechanism for the origin of the
extended light in our sample, although there is no evidence in general
for asymmetric scattering in radio-quiet QSOs.

%\subsection{Nebular light}
%
%  Nebular continuum produced by an extended
%narrow line region around the active nucleus is unlikely to be the
%origin of the hosts we have detected, the reason being that the
%expected narrow emission lines should be very prominent in that case,
%with peak intensities of the Balmer lines more than 3 times larger
%than those of the broad lines (ABT95). 
%The near-IR spectra of two of the QSOs studied here  show
%prominent broad \Ha\ lines, but no prominent narrow components
%(Aretxaga et al. 1997)

\subsection{Stellar light}

Stellar light remains still the most plausible interpretation for the
extended light we have detected. The colours, sizes, luminosities and
radial profiles are indeed in agreement with those expected from young
stellar populations:

a) The {\it colours} of the hosts, $R-K \approx 3.3$~mag for 1630.5$+$3749
and $R-K \lsim 3.3$ for PKS 2134$+$008 and Q 2244$-$0105, are not
consistent
with the colours predicted from a simple passively evolved stellar population,
usually assumed to be characteristic of
elliptical galaxies. 
A passively-evolved stellar population for a $z\sim2$ 
elliptical galaxy has colours  
$R-K \approx5.5$~mag and slightly bluer for a mild 
amount of activity [4].
Younger populations are necessary in order to 
account for the blue colours observed in our hosts. As an example,
typical nearby H~II galaxies set at $z=2$
would have colours as blue as $R-K \sim 1-1.7$,
as derived from their characteristic 
flat spectral energy distributions (SED): $f_\nu \propto \nu^\alpha$, 
with $0 \lsim \alpha \lsim 0.5$.

b) and c) The {\it luminosities and radii} of our hosts lie along the 
luminosity--radius relationship of the local young H~II galaxies 
[2].
And  there is
at least one local H~II galaxy that is as big and luminous as our hosts.

d) The {\it radial profiles} of 
the $R$ and $I$-band hosts, derived from the flat-top solutions, 
fall approximately as $r^{1/4}$-laws or exponential profiles for radii 
$r \gsim 0.6$~arcsec [2].
Profiles derived for radii smaller than the FWHM of the observations 
are usually unreliably 
recovered by flat-top subtractions, as shown by our numerical
simulations of galaxy$+$PSF. 

Therefore, the hosts of $z\approx2$ high-luminosity
radio-quiet and radio-loud QSOs are large and luminous, rivaling the
most luminous nearby star forming galaxies. Colours, sizes,
luminosities and radial profiles are compatible with the
interpretation of the extended light as a young galaxy.  If all the UV
luminosity is coming from a stellar population, the implied star
formation rates (in the continuous star formation case) are SFR$\gsim
100-200$~\Msun/yr on scales of $D_{eff}\approx 4$~Kpc. These 
values are about an order of magnitude higher than those derived for field
galaxies at similar redshifts found by Lyman Break searches in the
Hubble Deep Field.  At
$z\approx 2$, an unevolved $L_\star$ galaxy with SED typical of a
star-forming galaxy would appear to be about 3 mag fainter than the
hosts we have detected.  The density of high-luminosity QSOs ($M_B
\lsim -28$~mag) at redshifts between $1\lsim z \lsim 3$ is about
10 Gpc$^{-3}$.
This means that the density of large luminous
galaxies like the ones considered here should be about 8 Gpc$^{-3}$ at
$1<z<3$, taking into account the non-detection cases of extensions [1] and the
relative under-abundance of radio-loud QSOs, which form less than 10\%
of the QSO population.

\vspace*{0.2cm}
\noindent
{\it Acknowledgements:}
This work was supported in part by the `Formation and Evolution of
Galaxies' network set up by the European Commission under contract 
ERB FMRX-CT96-086 of its TMR programme. 

\vspace*{0.2cm}
\small
\noindent
{\bf References}

\noindent
 [1] Aretxaga I., Boyle B.J. \& Terlevich R.J. 1995, MNRAS, 275, L27 

\noindent
 [2] Aretxaga I., Terlevich R.J. \& Boyle B.J. 1997, MNRAS, in press (astro-ph/9711130)

\noindent
 [3] Bahcall J.N. et al. 1997, ApJ, 479, 642

\noindent
 [4] Bressan A. Chiossi C. \& Fagotto F. 1994, ApJS,94,63

\noindent
 [5] Cimatti A. et al. 1997,ApJ, 476, 677

\noindent
 [6] Hamman F. \& Ferland G. 1993, ApJ, 418, 11

\noindent
 [7] Isaak et al. 1994, MNRAS,269, 28

\noindent
 [8] Kauffmann G. \& Charlot S., 1997, MNRAS, in press (astro-ph/9704148)

\noindent
 [9] Lehnert M.D et al. 1992, ApJ, 393, 68

\noindent
 [10] Pritchet C.J. 1994, PASP, 106, 1052

\noindent
 [11] Terlevich R.J. \& Boyle, B.J., 1993, MNRAS, 262, 491

\end{document}